# Domain-stored skyrmion structures for a reading error-detectable racetrack memory

Tatsuro Karino[1]*, Daigo Shimizu[1], Ayaka P. Ohki[1], Nobuyuki Ikarashi[1,2], Takeshi Kato[1,2], Daiki Oshima[1]*, and Masahiro Nagao[1,2]*

[1] Department of Electronics, Nagoya University, Nagoya, Japan.

[2] Institute of Materials and Systems for Sustainability, Nagoya University, Nagoya, Japan.

**Abstract**

**Magnetic racetrack memory (RTM) uses a series of either domains or skyrmions as data bits along nanowires. However, it lacks reading error detection capability in nanowires, which requires high electric current density for the deterministic motion of domain walls (DWs) or skyrmions. Here, we propose a method of a reading error-detectable RTM and explore domain-stored skyrmion structures for this RTM. This method uses domains as memory cells and assigns the presence and absence of a skyrmion in a domain to different bits, enabling the electrical signal of any memory cells to output the opposite sign to that of the adjacent memory cells. Using simulations and Lorentz microscopy, we demonstrate that perpendicularly magnetized nanowires with relatively large Dzyaloshinskii-Moriya interaction can achieve domain-stored skyrmion structures. Additionally, our simulations show that when DWs approach to skyrmions during spin-orbit torque-induced motion, angular momentum transfers from DWs to skyrmions, resulting in fast motion of skyrmions.**

## Introduction

Magnetic racetrack memory (RTM) is a non-volatile solid-state memory that uses current-induced spin torque to move domain walls (DWs) or skyrmions bi-directionally to the read and write elements in individual nanowires[1,2] (Fig. 1a,b, top and middle panels). The data is encoded in a series of magnetic structures where the upward or downward magnetization direction of a domain or the presence or absence of a skyrmion represents a single bit of binary data. Perpendicularly magnetized multilayer nanowires allow spin-orbit torque (SOT) to drive Néel-type chiral DWs or skyrmions stabilized by the interfacial Dzyaloshinskii-Moriya interaction (DMI) at high efficiency and speed[3,4]. This makes it an appealing way for low power consumption and high-speed processing in RTM.

RTM requires precise position control of the spin textures along nanowires. In regions where the same data bits are adjacent, the same magnetic structure aligns, resulting in the output of the identical electrical signal, as depicted in the bottom panels of Fig. 1a,b. However, relying solely on electric current pulses can cause an unexpected misalignment of the spin textures due to their stochastic motions originating from natural pinning sites[5,6,7]. This can lead to reading errors (in the case of skyrmions, at the region where they are absent). To address this issue, researchers have introduced artificial pinning sites such as notches[1,8,9,10,11], local modulation of magnetic properties[12,13,14,15,16,17,18], and electrostatic potential[19,20]. While these approaches can achieve highly reliable motion, they require a significantly high current density to deterministically depin the spin textures from natural and artificial pinning sites. The need for high current density arises from a lack of position error detection capability in individual nanowires. Here, we propose a method of a reading error-detectable RTM that exploits both Néel-type chiral DWs and skyrmions. The magnetic structures are that zero-field skyrmions exist between DWs in perpendicularly magnetized multilayer nanowires. We hereafter refer to the structures as "domain-stored skyrmion structures". Our simulations and Lorentz transmission electron microscopy (L-TEM) observations demonstrate that such structures are feasible in nanowires with relatively high DMI. Additionally, our simulations show that when DWs approach to skyrmions, angular momentum transfers from DWs to skyrmions, resulting in fast motion of skyrmions, as if more highly efficient SOT-induced motion.

## Method of a reading error-detectable RTM

The proposed reading error-detectable RTM uses both DWs and skyrmions, different from conventional RTMs that use either DWs or skyrmions. The method uses domain-stored skyrmion structures in the perpendicularly magnetized nanowires, as depicted in the top panel of Fig. 1c. (Although, spin textures with a topological charge of $Q = -1$ and $Q = 1$ in the upward and downward domains are defined as a skyrmion and an antiskyrmion, respectively, this Article collectively describes both as the skyrmion for convenience.) In conventional RTMs, there are no magnetic boundaries between the adjacent same data bits (except for the region of adjacent skyrmions). On the other hand, our method uses individual domains as memory cells and assigns the presence and absence of the skyrmion in a domain to two different bits (Fig. 1c, top and middle panels). This method allows the boundaries (DWs) to always exist between memory cells, even when the same bits are adjacent, and the electrical signal of any memory cells can always output the opposite sign to that of the adjacent memory cells (Fig. 1c, bottom panel), enabling the detection of reading errors in individual nanowires. While RTM with both skyrmions and anisotropic antiskyrmions may also be capable of reading error detection[21], experiments have reported only collective transformation between antiskyrmions and Bloch-type skyrmions using in-plane magnetic fields[21,22,23]. Domain-stored skyrmion structure can adopt state-of-art application technologies[24,25,26,27,28,29] and thus is a promising magnetic structure for the reading error-detectable RTM. In the following, we explore the feasibility of domain-stored skyrmion structures in perpendicularly magnetized multilayer nanowires.

## Micromagnetic simulations of skyrmions in multi-domain structures

A fundamental impediment to realizing the reading error-detectable RTM is the absence of domain-stored skyrmion structures in perpendicularly magnetized multilayer nanowires. There are also no detailed studies on magnetic structures where zero-field skyrmions are contained in multi-domain structures in multilayer thin films. We hereafter refer to the structures in thin films as "domain-contained skyrmion structures". Using micromagnetic simulations, we create magnetic phase diagrams to determine the realistic magnetic parameter regions for domain-contained skyrmion structures. Figure 2 shows the universal zero-field $D$-$K_u$ phase diagrams at $A$ = 5, 10, and 15 pJ/m in thin films of the thickness of $t$ = 1 nm, where $D$, $K_u$, and $A$ are the DMI, perpendicular magnetic anisotropy, and exchange stiffness energy, respectively. Our simulations suppose the Co-based multilayers, which are typical materials hosting skyrmions[30,31,32]. The phase diagrams depict each panel as a magnetization distribution map. The black and white colour regions correspond to downward and upward out-of-plane magnetization directions, respectively. We find that domain-contained skyrmions and similar structures (skyrmion bags[33,34]) are present in relatively high $D$ values at every $A$. The cyan lines represent the magnetic structural boundaries between the presence and absence of skyrmions. Some magnetic structures have skyrmions in the monodomain, but we observe the presence of domain-contained skyrmions and skyrmion bags in a larger calculated area. The magenta lines represent the energy boundaries between $E_{S-D} < E_{FM}$ and $E_{S-D} > E_{FM}$, where $E_{S-D}$ and $E_{FM}$ are the relaxed micromagnetic energy and the uniform ferromagnetic energy, respectively. We focus on the size of skyrmions and domains. At every $A$, a decrease in the skyrmion size as a function of $D$ and $K_u$ is consistent with that in a previous study[2]. When $K_u$ increases with fixed $D$, the skyrmion size decreases at the metastable states with higher energy in the region of $E_{S-D} > E_{FM}$.

When $D$ decreases with fixed $K_u$, skyrmions show the same behaviour. However, in some panels, we note a discontinuous increase in the size of skyrmions (for example, $A = 10$ pJ/m, $D = 1.2$ mJ m$^{-2}$, $K_u = 0.35$ MJ m$^{-3}$). These exceptions could be due to competition with the ferromagnetic state (see Supplementary Fig. S1). An increase in the domain size follows a general trend. As $K_u$ ($D$) increases (decreases), small irregularly-shaped domains alter to large circular bags. The position and slope of the cyan and magenta lines depend on $A$, which can be understood by the skyrmion stability[35] (see Supplementary Note 1). These results show that multilayer thin films can form domain-contained skyrmion structures at realistic magnetic parameter values and are a basis for obtaining domain-stored skyrmion structures in nanowires.

**Micromagnetic simulations of domain-stored skyrmion structures**

We determine the magnetic parameter regions that allow for domain-stored skyrmion structures in nanowires based on the phase diagrams of thin films. Figure 3a shows the evolution of the magnetic structure as a function of wire width $w$ at $A = 10$ pJ/m, $D = 1.2$ mJ m$^{-2}$, and $K_u = 0.45$ MJ m$^{-3}$ (green circle in Fig. 2). At $w = 512$ nm, DWs form along the short axis of the wire, while skyrmions remain, and half-skyrmion bags appear at the edges. As the wire gets narrower ($w = 256$ nm), the repulsion between the periphery of the half-skyrmion bags and the opposite edges destabilizes the half-skyrmion bags and consequently annihilates them, leading to the realization of a domain-stored skyrmion structure.

Domain-stored skyrmion structures in nanowires do not appear in the entire region where domain-contained skyrmion structures appear in thin films. Figure 3b is the same as Fig. 3a, but only $K_u$ is changed to 0.35 MJ m$^{-3}$ (red circle in Fig. 2). At $w = 256$ nm, the magnetic structure differs little from that in the thin film. At $w = 128$ nm, many half-bags without skyrmions and only one skyrmion appear at the edges and the center, respectively. When the wire gets narrower ($w = 64$ nm), the skyrmion vanishes, and the half-bag remains.

We identify the formation regions in nanowires by developing a colour-coded phase diagram at $A = 10$ pJ/m (Fig. 3c). The cyan and magenta lines are the same as in Fig. 2 at $A = 10$ pJ/m. Supplementary Fig. S2 displays the colour-coded phase diagrams at $A = 5$ and 15 pJ/m. Our analysis reveals that the formation regions (green panels) are located only in the $E_{\text{S-D}} < E_{\text{FM}}$ region at $A = 10$ pJ/m. This pattern also is observed at $A = 5$ and 15 pJ/m. Based on Fig. 2 and 3a,b,c, we conclude that domain-stored skyrmion structures in nanowires can form when skyrmions are sufficiently smaller than domains in thin films. The boundaries between the formation and non-formation regions roughly coincide with the magenta lines, providing further evidence to support our conclusion. We provide additional information in Supplementary Table S1.

We confirm that skyrmions exhibit SOT-induced motion along nanowires while stored in domains. Figure 3d presents the time-dependent dynamics in a part of the nanowire. We describe the dynamics within the orange boxes in Fig. 3d. On applying a current density of $j_{\text{SOT}} = 2.0 \times 10^{10}$ A m$^{-2}$ along the nanowire ($t = 3$ ns), the DW and skyrmion begin to move in the same direction. The skyrmion moves with the transverse component, known as the skyrmion Hall effect. The DW gradually shows tilt induced by the DMI[36]. The DW moves longitudinally faster than the skyrmion, resulting in distortion in close proximity to the skyrmion ($t = 5$ ns). The dotted line represents the DW tilting in the absence of the skyrmion. Interestingly, from $t = 5$ to 6 ns, the upper segment of the DW stops while the skyrmion gains speed and then in steady-state motion at high speed. The skyrmion Hall angle during this event is the same as $\Theta_{\text{Sk}} \approx 78°$ of isolated skyrmions. When the skyrmion reaches the edges, the upper segment moves again at high speed, pushing the skyrmion forward ($t = 15$ ns). Finally, the domain regains the shape, and the DW and skyrmion show steady-state motion with the same velocity ($t = 20$ ns). The interactive movements between the DW and skyrmion arise from angular momentum transfer, as we discuss below. Note that above $j_{\text{SOT}} = 3.0 \times 10^{10}$ A m$^{-2}$, some skyrmions disappear as a result of strong interaction with

DWs.

## Observation of domain-stored skyrmion structure

We also conduct experiments to strengthen the feasibility of domain-stored skyrmion structures. We prepare Pt/Co/Ta multilayer thin films by magnetron sputtering. Figure 4a displays the in-plane and out-of-plane magnetization loops in [Pt(1.3 nm)/Co(1.0 nm)/Ta(0.7 nm)]$_{15}$ thin films grown at an Ar gas pressure of $P_{Ar}$ = 0.4 Pa, which is a promising sample among the ones we prepared, with $K_{eff}$ = 0.50 MJ m$^{-3}$ and $M_s$ = 807 kA/m (see Supplementary Fig. S3 and Table S2). The former corresponds to $K_u \approx$ 0.75 MJ m$^{-3}$ in Fig. 2, and the latter is roughly half the value of bulk Co due to the presence of the dead layers[37] and slightly larger than the value used in our simulations. In this as-grown thin film, at zero field, the L-TEM image at a sample tilt angle of $\theta$ = 0° shows almost no contrast (Supplementary Fig. S4), but do the bright and dark line contrasts at $\theta$ = 30°, indicating the Néel-type DW structure[38] (Fig. 4b). However, this sample forms the multi-domain structure without skyrmions, which could be due to the insufficient $D$. To obtain a multilayer with higher $D$, we prepare another sample with the same configuration but at $P_{Ar}$ = 2.0 Pa (2.0-sample). Our recent study has confirmed the enhancement of DMI through this method using the high Ar gas pressure, and defects are likely to be responsible for this enhancement[39]. The magnetization loops in the 2.0-sample are analogous to those in the 0.4-sample (Fig. 4c), but the L-TEM image at $\theta$ = 30° shows that the magnetic structure changes, with domain-contained skyrmions in the as-grown 2.0-sample (Fig. 4d). Note that the use of high Ar gas pressures harms the crystal quality[40], leading to reduction of $A$.

We now explore domain-stored skyrmions in nanowires. The previous 2.0-sample has dense skyrmions (Fig. 4d and Supplementary Fig. S5), suggesting equilibrium skyrmions unsuitable for the application. To obtain a metastable skyrmion, we prepared a sample with Pt thickened by 0.2 nm from the previous 2.0-sample for increasing $K_u$, according to the findings in Fig. 2. As a result, we observed sparse skyrmions in multi-domain structure at zero field (Supplementary Fig. S5). We fabricated the nanowire with $w \approx$ 700 nm of this 2.0-sample (see Methods) and then conducted L-TEM observations. Figure 4e-g is a series of the L-TEM images as a function of out-of-plane magnetic fields $B$. Out-of-plane magnetization loop is shown in Fig. 4h. At zero field after applying $B$ = −2 T, there is a monodomain state without any skyrmions (Fig. 4e). At $B$ = +40 mT, three skyrmions and three DWs appear (Fig. 4f). On the removal of the magnetic field, every DW and skyrmion remains in approximately the same position as at $B$ = +40 mT (Fig. 4g). The L-TEM images show contrast inversion by tilting the sample in the inverse direction (Fig. 4i,j), which provides further confirmation that the contrasts originate from DWs and skyrmions. (The DW tilting is due to a weak residual magnetic field in TEM.) The results indicate that multilayer nanowires can achieve domain-stored skyrmion structures. We also observe a few DWs nearly parallel to the wire longitudinal direction (Supplementary Fig. S6), indicating an incomplete domain-stored skyrmion structure due to wider than adequate wire widths. Although our L-TEM makes it hard to observe unambiguous magnetic contrasts in nanowires below $w$ = 700 nm (see Methods), observing partially domain-stored skyrmions suggests that complete structures appear in narrower wires. We also stress that high-quality materials with high $D$ are necessary for applications because this sample is weak against heating due to reduced crystal quality and can thus not withstand Joule heating. Since the research on the interfacial DMI is progressing rapidly, we expect to obtain such materials in the near future. Alternatively, domain-stored skyrmion structures could be achievable in synthetic antiferromagnets[41] in which the thickness of two ferromagnetic layers is slightly different, and ferrimagnets[42]. They can suppress the skyrmion Hall effect and DW tilting and realize small skyrmions because of vanishingly dipolar interaction.

**Angular momentum transfer from DWs to skyrmions**

Finally, we discuss the interactive movement between DWs and skyrmions. In particular, we focus on the transfer of angular momentum[43]. The transfer exhibits unique features, as demonstrated by the density maps of total torque ($\tau$) (Fig. 5a, and see Supplementary Fig. S7 for resolving Fig. 5a into $x$-, $y$-, and $z$-components). Figure 5b shows the time-resolved longitudinal and transverse skyrmion velocities. We describe the dynamics within the orange boxes in Fig. 5a. Upon injection of a current, a finite torque appears at the DW and skyrmion ($t$ = 2.5 ns), with the torque on the DW ($\tau_{DW}$) being more magnitude than the skyrmion ($\tau_{Sk}$). The DW concentrates $\tau_{DW}$ in its upper region to gain speed. This behaviour is because the stable geometrical shape of the moving DW is a tilted structure[36]. When the DW approach to the skyrmion ($t$ = 6.5 ns), $\tau_{DW}$ decreases in the distorted region of the DW, while $\tau_{Sk}$ increases. At 13.0 ns, $\tau_{DW}$ vanishes in the upper segment of the DW, while $\tau_{Sk}$ further increases. As a result, the segment stops while the skyrmion gains speed. The map at 20.0 ns has almost the same torque intensity as that at 13.0 ns. At 20.0 ns, the skyrmion is in steady-state motion at high speed while the upper segment remains at rest. These changes in the torque distribution indicate that the DW transfers angular momentum to the skyrmion, which we can understand in terms of DW stability. At $t$ = 13.0-20 ns, the DW shape deviates from the stable tilted structure (white dotted lines) due to the presence of the skyrmion. Hence, the DW attempts to transition to a tilt-free structure by stopping, leading to leakage of angular momentum into the skyrmion. As the skyrmion reaches the edge and begins to leave the DW ($t$ = 30.0-35.0 ns), $\tau_{DW}$ increases in the upper segment, similar to that at $t$ = 2.5 ns. Simultaneously, $\tau_{Sk}$ also increases, resulting in the skyrmion gaining speed along the edge. As the DW gradually regains shape, $\tau_{DW}$ distribution shows bias reduction, and $\tau_{Sk}$ decreases. Finally, $\tau$ shows no time dependence, resulting in the steady-state motion. The skyrmion (DW) moves at a speed 1.84 (1.04) times faster (slower) than in the absence of the DW (skyrmion) (Supplementary Fig. S8), indicating a steady leakage of angular momentum from the DW to the skyrmion. At $t$ = 30.0-40.0 ns, the increase in both $\tau_{DW}$ and $\tau_{Sk}$ is peculiar, which could relate to angular momentum transfer from the edge. Indeed, our simulations show that, in the case of an isolated skyrmion, $\tau_{Sk}$ increases when the skyrmion reaches the edge (Supplementary Fig. S9). This increase could be due to a torque acting on the chiral edge twist at the edge[44]. Our results highlight the unique dynamics of the coexisting state of DWs and skyrmions. Nonetheless, experimental and detailed theoretical studies are necessary to fully understand these angular momentum transfers.

**Conclusions**

We have presented a method of the reading error-detectable RTM that uses domain-stored skyrmion structures in nanowires. Our simulations indicate that domain-stored skyrmion structures can be achieved in perpendicularly magnetized multilayer nanowires with a relatively large DMI and sufficiently smaller skyrmions than domains. These structures form under a metastable state, making them ideal for practical applications. In addition, L-TEM observations confirm the feasibility of such structures. Although the proposed RTM still requires artificial pinning sites, it could reduce excessive current and lower redundancy levels, leading to high efficiency and storage density RTM. Our simulations also show angular momentum transfer from DWs to skyrmions, suggesting that skyrmions can gain angular momentum from other spin textures subjected to SOT. This finding provides new insight into angular momentum transfer of spin textures, which may help to develop more efficient manipulation of SOT-driven skyrmion motion and design of spintronics devices.

## Methods

Micromagnetic simulations

We conducted micromagnetic simulations using Mumax$^3$ (ref. 45). Magnetic structures are simulated by relaxation from a random state of magnetization through a conjugate gradient method with a size of 1024 × 1024 × 1 nm$^3$ on a 512 × 512 × 1 mesh for thin films (Fig. 2, and 3a,b) and with 4096 × $X$ × 1 nm$^3$ where $X$ is 64, 128, 256, 512, 1024, 2048, and 4096 nm using 2 × 2 × 1 nm$^3$ cell size for nanowires (Fig. 3c,d). The continuum energy function contains the exchange stiffness, DMI, demagnetizing field, and perpendicular magnetic anisotropy energy. For thin films, the periodic boundary conditions are applied in the $x$- and $y$-directions for thin films. For nanowires, the periodic boundary condition is applied in the $x$-directions and the free boundary condition is applied in the $y$-directions. In this study, we suppose Co-based multilayers and thus used a fixed value of the saturation magnetization $M_s$ = 700 kA/m (ref. 37). Current-driven dynamic simulations were conducted by micromagnetic simulations based on the Landau-Lifshitz-Gilbert (LLG) equation. The simulations used $A$ = 10 pJ/m, $D$ = 1.2 mJ m$^{-2}$, $K_u$ = 0.45 MJ m$^{-3}$ and a Gilbert damping of $\alpha$ = 0.16 (ref. 46). SOTs were implemented by adding the torques to the LLG equation$^{47}$. The spin Hall angle is $\alpha_H$ = 0.07. The field-like parameter, which is the ratio of the field-like and damping-like torque, is $\xi$ = 0. We have also simulated with $\xi$ = −2 and 5, and the results are same as to the case with $\xi$ = 0. Therefore, in the main text, we only present the results at $\xi$ = 0.

Experiments

Multilayer thin films of Ta(5 nm)/[Pt($t_{Pt}$ nm)/Co($t_{Co}$ nm)/Ta($t_{Ta}$ nm)]$_N$/Pt(3 nm) were grown using d.c. magnetron sputter deposition at room temperature under an Ar pressure of 0.2, 0.4 and 2.0 Pa. The top layer with 3-nm-thick Pt is the protective layer against oxidation and the bottom layer with 5-nm-thick Ta is the buffer layer. $N$ is the number of repetition layers. We prepared 44 samples (see Supplementary Table S2). We deposited the thin films on 10-nm and 15-nm-thick Si$_3$N$_4$ membranes used for L-TEM imaging, and on thermally oxidized Si wafers used for vibrating sample magnetometry measurement. The thin films on two different substrates were grown simultaneously. We aimed for large effective perpendicular magnetic anisotropy $K_{eff}$ while paying attention to the saturation magnetization $M_s$ value and number of the repetition $N$ because there is a trade-off between increasing L-TEM contrast intensity and suppressing dipolar interaction. The samples displayed in Fig. 4 have relatively low $M_s$ and relatively high $K_{eff}$ among the samples that we prepared (Supplementary Fig. S3). The nanowires were fabricated by a focused ion beam technology. To suppress Fresnel fringes from the wire edges, using focused Ga$^+$ beam, we made two parallel trenches in the thin film until the Si$_3$N$_4$ membrane was exposed, and depositing amorphous Pt in two trenches.

Observations of magnetic structures were performed using the Fresnel mode of L-TEM (JEM2100F, JEOL) with the double-tilting holder where the maximum angle of tilt is $\theta$ = ± 30°. To avoid the artificial distortion of the magnetic contrasts, the astigmatism was corrected using magnification calibration diffraction grating replica. The value of the magnetic field was changed by controlling the current of the objective lens. Our L-TEM experiment set a relatively large defocus that is enough not to superimpose the compelling Fresnel fringes on the magnetic contrasts because of the small $M_s$. Consequently, the L-TEM observation is difficult for nanowires below $w$ = 700 nm.

## Data availability

The data generated during and/or analysed during the current study are available from the corresponding author(s) on reasonable request.

## References

1. Parkin, S. S. P., Hayashi, M. & Thomas, L. Magnetic domain-wall racetrack memory. *Science* **320**, 190–194 (2008).
2. Sampaio, J., Cros, V., Rohart, S., Thiaville, A. & Fert, A. Nucleation, stability and current-induced motion of isolated magnetic skyrmions in nanostructures. *Nat. Nanotechnol.* **8**, 839–844 (2013).
3. Ryu, K-S., Thomas, L., Yang, S-H. & Parkin, S. S. P. Chiral spin torque at magnetic domain walls. *Nat. Nanotechnol.* **8**, 527–533 (2013).
4. Emori, S., Bauer, U., Ahn, S-M., Martinez, E. & Beach, G. S. D. Current-driven dynamics of chiral ferromagnetic domain walls. *Nat. Mater.* **12**, 611–616 (2013).
5. Lemerle, S., Ferré, J., Chappert, C., Mathet, V., Giamarchi, T. & Le Doussal, P. Domain wall creep in an Ising ultrathin magnetic film. *Phys. Rev. Lett.* **80**, 849–852 (1998).
6. Chauve, P., Giamarchi, T. & Le Doussal, P. Creep and depinning in disordered media. *Phys. Rev. B* **62**, 6241–6267 (2000).
7. Meier, G., Bolte, M., Eiselt, R., Krüger, B, Kim, D-H., & Fischer, P. Direct imaging of stochastic domain-wall motion driven by nanosecond current pulses. *Phys. Rev. Lett.* **98**, 187202 (2007).
8. Hayashi, M., Thomas, L., Rettner, C., Moriya, R., Jiang, X., & Parkin, S. S. P. Dependence of current and field driven depinning of domain walls on their structure and chirality in permalloy nanowires. *Phys. Rev. Lett.* **97**, 207205 (2006).
9. Atkinson, D., Eastwood, D. S. & Bogart, L. K. Controlling domain wall pinning in planar nanowires by selecting domain wall type and its application in a memory concept. *Appl. Phys. Lett.* **92**, 585 (2008).
10. Huang, S-H. & Lai, C-H. Domain-wall depinning by controlling its configuration at notch. *Appl. Phys. Lett.* **95**, 032505 (2009).
11. Yuan, H. Y. & Wang, X. R. Domain wall pinning in notched nanowires. *Phys. Rev. B* **89**, 054423 (2014).
12. Ummelen, F., Swagten, H. & Koopmans, B. Racetrack memory based on in-plane-field controlled domain-wall pinning. *Sci. Rep.* **7**, 833 (2017).
13. Migita, K., Yamada, K., & Nakatani, Y. Controlling skyrmion motion in an angelfish-type racetrack memory by an AC magnetic field. *Appl. Phys. Exp.* **13**, 073003 (2020).
14. Polenciuc, I., Vick, A. J., Allwood, D. A., Hayward, T. J., Vallejo-Fernandez, G., O'Grady, K. & Hirohata, A. Domain wall pinning for racetrack memory using exchange bias. *Appl. Phys. Lett.* **105**, 162406 (2014).
15. Yoon, J., Yang, S-H., Jeon, J-C., Migliorini, A., Kostanovskiy, I., Ma, T. & Parkin, S. S. P. Local and global energy barriers for chiral domain walls in synthetic antiferromagnet–ferromagnet lateral junctions. *Nat. Nanotechnol.* **17**, 1183–1191 (2022).
16. Rantschler, J. O., McMichael, R. D., Castillo, A., Shapiro, A. J., Egelhoff, W. F., Maranville, B. B., Pulugurtha, D., Chen, A. P. & Connors, L. M. Effect of 3d, 4d, and 5d transition metal doping on damping in permalloy thin films. *J. Appl. Phys.* **101**, 033911 (2007).
17. Lee, T., Jeong, S., Kim, S. & Kim, K-J. Position-reconfigurable pinning for magnetic domain wall motion. *Sci. Rep.* **13**, 6791 (2023).
18. Wu, H. Z., Miao, B. F., Sun, L., Wu, D. & Ding, H. F. Hybrid magnetic skyrmion. *Phys. Rev. B* **95**, 174416 (2017).
19. Bauer, U., Emori, S. & Beach, G. S. D. Voltage-controlled domain wall traps in ferromagnetic nanowires. *Nat. Nanotechnol.* **8**, 411–416 (2013).
20. Franke, K. J. A., Wiele, B. V., Shirahata, Y., Hämäläinen, S. J., Taniyama, T. & Dijken, S. Reversible electric-field-

driven magnetic domain-wall motion. *Phys. Rev. X* **5**, 011010 (2015).

21. Jena, J., Göbel, B., Kumar, V., Mertig, I., Felser, C. & Parkin, S. Evolution and competition between chiral spin textures in nanostripes with $D_{2d}$ symmetry. *Sci. Adv.* **6**, eabc0723 (2020).

22. Peng, L. C. et al. Controlled transformation of skyrmions and antiskyrmions in a non-centrosymmetric magnet. *Nat. Nanotechnol.* **15**, 181–186 (2020).

23. Karube, K. et al. Room-temperature antiskyrmions and sawtooth surface textures in a non-centrosymmetric magnet with $S_4$ symmetry. *Nat. Mater.* **20**, 335–340 (2021).

24. Hanneken, C., Otte, F., Kubetzka, A., Dupé, B., Romming, N., von Bergmann, K., Wiesendanger, R. & Heinze, S. Electrical detection of magnetic skyrmions by tunnelling non-collinear magnetoresistance. *Nat. Nanotechnol.* **10**, 1039–1042 (2015).

25. Maccariello, D., Legrand, W., Reyren, N., Garcia, K., Bouzehouane, K., Collin, S., Cros, V. & Fert, A. Electrical detection of single magnetic skyrmions in metallic multilayers at room temperature. *Nat. Nanotechnol.* **13**, 233–237 (2018).

26. Penthorn, N. E., Hao, X., Wang, Z., Huai, Y. & Jiang, H. W. Experimental observation of single skyrmion signatures in a magnetic tunnel junction. *Phys. Rev. Lett.* **122**, 257201 (2019).

27. Büttner, F., Lemesh, I., Schneider, M., Pfau, B., Günther, C. M., Hessing, P., Geilhufe, J., Caretta, L., Engel, D., Krüger, B., Viefhaus, J., Eisebitt, S. & Beach, G. S. D. Field-free deterministic ultrafast creation of magnetic skyrmions by spin–orbit torques. *Nat. Nanotechnol.* **12**, 1040–1044 (2017).

28. Woo, S., Song, K. M., Zhang, X. C., Ezawa, M., Zhou, Y., Liu, X. X., Weigand, M., Finizio, S., Raabe, J., Park, M. C., Lee, K. Y., Choi, J. W., Min, B. C., Koo, H. C. & Chang, J. Deterministic creation and deletion of a single magnetic skyrmion observed by direct time-resolved X-ray microscopy. *Nat. Electron.* **1**, 288–296 (2018).

29. Chen, S. et al. All-electrical skyrmionic magnetic tunnel junction. *Nature* **627**, 522–527 (2024).

30. Moreau-Luchaire, C. et al. Additive interfacial chiral interaction in multilayers for stabilization of small individual skyrmions at room temperature. *Nat. Nanotechnol.* **11**, 444–448 (2016).

31. Woo, S. et al. Observation of room-temperature magnetic skyrmion and their current-driven dynamics in ultrathin metallic ferromagnets. *Nat. Mater.* **15**, 501–506 (2016).

32. Wang, L. et al. Construction of a room-temperature Pt/Co/Ta multilayer film with ultrahigh-density skyrmions for memory application. *ACS Appl. Mater. Interfaces* **11**, 12098–12104 (2019).

33. Zeng, Z. et al. Dynamics of skyrmion bags driven by the spin–orbit torque. *Appl. Phys. Lett.* **117**, 172404 (2020).

34. Foster, D. et al. Two-dimensional skyrmion bags in liquid crystals and ferromagnets. *Nat. Phys.* **15**, 655–659 (2019).

35. Bogdanov, A. & Hubert, A. Thermodynamically stable magnetic vortex states in magnetic crystals. *J. Magn. Magn. Mater.* **138**, 255–269 (1994).

36. Boulle, O., Rohart, S., Buda-Prejbeanu, L. D., Jué, E., Miron, I. M., Pizzini, S., Vogel, J., Gaudin, G. & Thiaville, A. Domain wall tilting in the presence of the Dzyaloshinskii-Moriya interaction in out-of-plane magnetized magnetic nanotracks. *Phys. Rev. Lett.* **111**, 217203 (2013).

37. Woo, S., Mann, M., Tan, A. J., Caretta, L. & Beach, G. S. D. Enhanced spin–orbit torques in Pt/Co/Ta heterostructures. *Appl. Phys. Lett.* **105**, 212404 (2014).

38. Pollard, S. D. et al. Observation of stable Néel skyrmions in cobalt/palladium multilayers with Lorentz transmission electron microscopy. *Nat. Commun.* **8**, 14761 (2017).

39. Ohki, A. P, Karino, T., Shimizu, D., Ikarashi, N., Kato, T., Oshima, D. & Nagao, M. Defect-mediated bulk Dzyaloshinskii-Moriya interaction in ferromagnetic multilayer thin films. arXiv
40. Messier, R., Giri, A. P. & Roy, R. A. Revised structure zone model for thin film physical structure. *J. Vac. Sci. Technol. A* **2**, 500–503 (1984).
41. Legrand, W. et al. Room-temperature stabilization of antiferromagnetic skyrmions in synthetic antiferromagnets. *Nat. Mater.* **19**, 34–42 (2020).
42. Hirata, Y. et al. Vanishing skyrmion Hall effect at the angular momentum compensation temperature of a ferrimagnet. *Nat. Nanotechnol.* **14**, 232–236 (2019).
43. Yang, S.-H., Garg, C. & Parkin, S. S. P. Chiral exchange drag and chirality oscillations in synthetic antiferromagnets. *Nat. Phys.* **15**, 543–548 (2019).
44. Rohart, S. & Thiaville, A. Skyrmion confinement in ultrathin film nanostructures in the presence of Dzyaloshinskii–Moriya interaction. *Phys. Rev. B* **88**, 184422 (2013).
45. Vansteenkiste, A. et al. The design and verification of MuMax3. *AIP Adv.* **4**, 107133 (2014).
46. Wang, Z. et al. Thermal generation, manipulation and thermoelectric detection of skyrmions. *Nat. Electron.* **3**, 672–679 (2020).
47. Litzius, K. et al. Skyrmion Hall effect revealed by direct time-resolved X-ray microscopy. *Nat. Phys.* **13**, 170–175 (2017).

## Acknowledgements

We thank Y. Yamamoto, K. Higuchi, and H. Cheong for technical support of experiments. This work was financially supported by Grant-in-Aid for Scientific Research (B) (JSPS, 21H01029). This work was conducted at Microstructure Analysis Platform and Microstructure Analysis Platform in the Next-generation biomaterials Hub, Nagoya University, supported by "Advanced Research Infrastructure for Materials and Nanotechnology in Japan (ARIM)" of the Ministry of Education, Culture, Sports, Science and Technology (MEXT).

## Contributions

T. Karino, and M.N. conceived and designed the research, and wrote the manuscript. T. Karino carried out the simulations and all the experiments. D.S performed the simulations. D.O., and T. Kato prepared the samples. A.O carried out L-TEM experiments. N.I. discussed the L-TEM data. All authors discussed the results and commented on the manuscript. D.O., and M.N. supervised the research.

## Corresponding authors

Correspondence to Tatsuro Karino, Daiki Oshima, or Masahiro Nagao.

## Ethics declarations

Competing interests

The authors declare no competing interests.

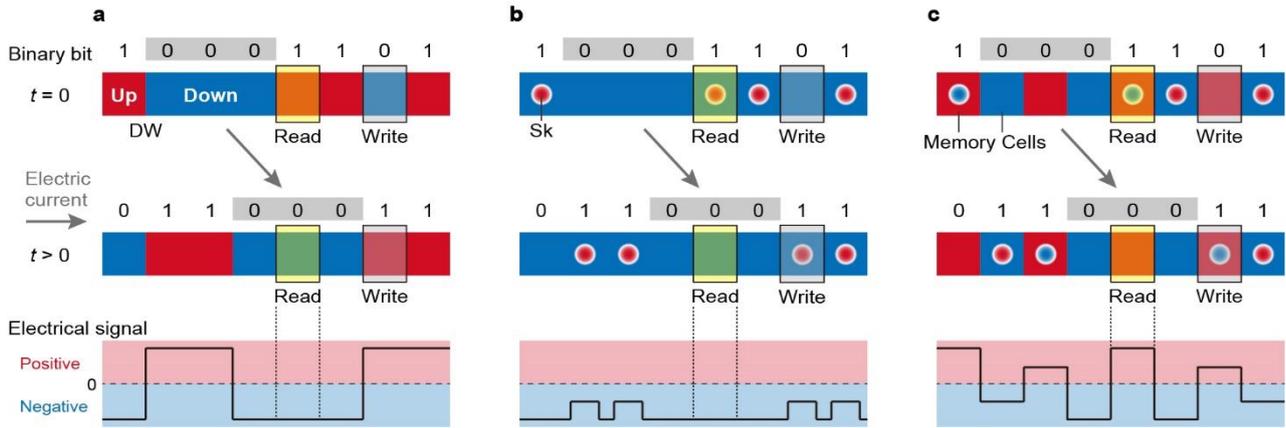

**Fig. 1 | Schematic of three RTMs and proposal for a reading error-detectable RTM. a,b,c,** Different types of three RTMs. DW- (**a**), skyrmion- (Sk-) (**b**), and domain-stored skyrmion-based RTMs (**c**). The last is the RTM proposed in this study. The top and middle panels represent magnetic configurations before and after applying an electric current $J_{SOT}$ for SOT generation, respectively. Red and blue regions represent upward and downward domains, respectively. Red and blue circles bordered by white lines represent skyrmions. 0 and 1 listed above the wires are bits. The gray shadows on the numbers highlight that the same data bits are adjacent. The bottom panels represent the sign switching of the electrical signal obtained from the magnetotransport measurement, corresponding to the magnetic states of the middle panels. In **c**, the sign of the electrical signal is alternately switched for a series of memory cells (domains).

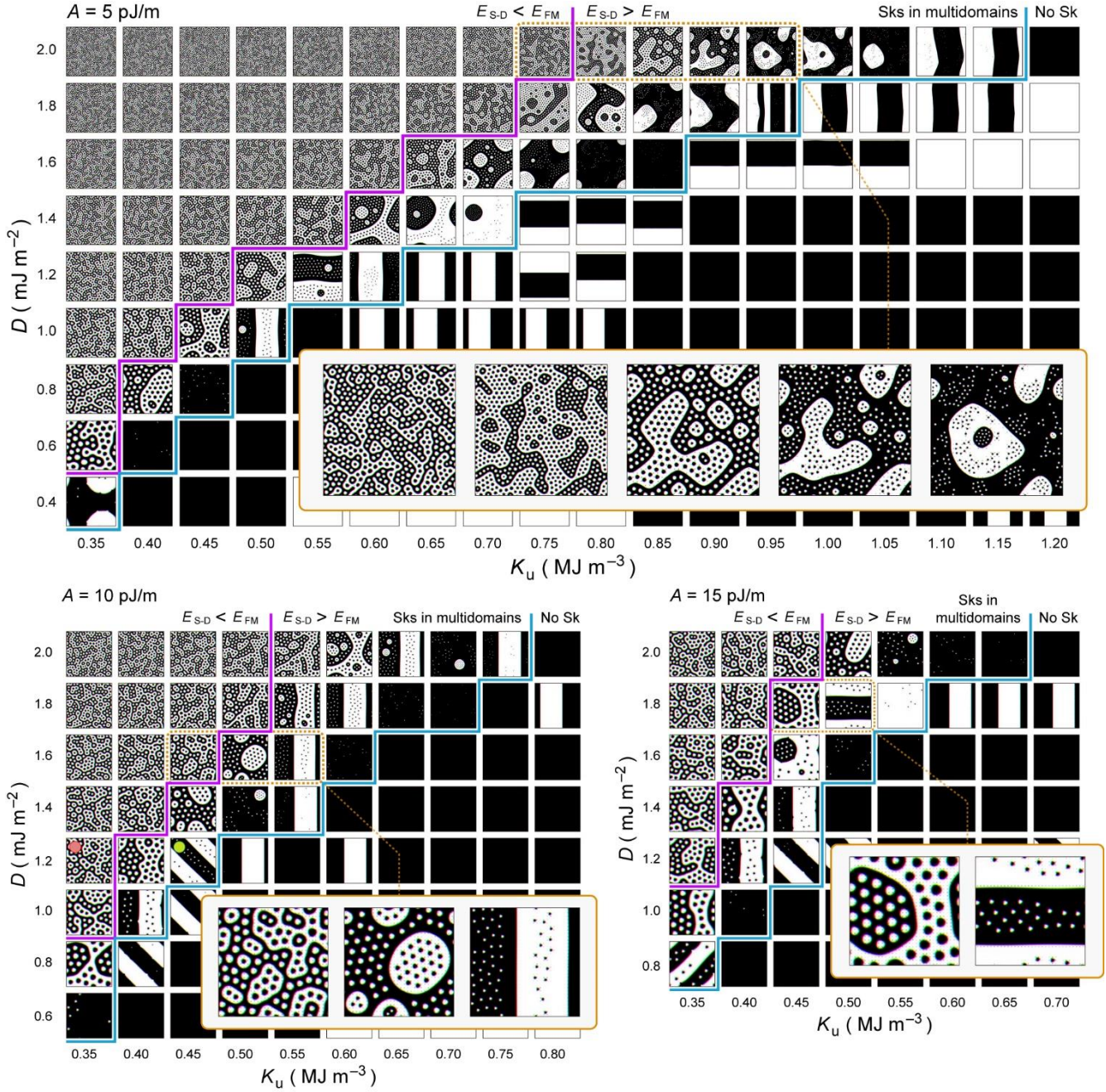

**Fig. 2 | Phase diagrams for coexisting of DWs and skyrmions.** Zero-field $D$-$K_u$ phase diagrams for $A$ = 5 (Top), 10 (bottom left), and 15 pJ/m (bottom right) in the thin film with $t$ = 1 nm. The phase diagrams depict each panel as a magnetization distribution map. The saturation magnetization[36] is $M_s$ = 700 kA/m. The black and white colour regions correspond to downward ($-m_z$) and upward ($+m_z$) out-of-plane magnetization directions, respectively. The upper limit of $D$ was set to 2.0 mJ m$^{-2}$ which is expected to be realized in the future[29]. The lower limit of $K_u$ was set to the minimum value that preserves effective perpendicular magnetic anisotropy $K_{eff}$. We draw the cyan lines as the structural boundaries between the presence (upper and left side) and absence (lower and right side) of skyrmions, and the magenta lines as the energy boundaries between $E_{S-D} < E_{FM}$ (upper and left side) and $E_{S-D} > E_{FM}$ (lower and right side). Note that we ignore the states where only a few skyrmions are formed. Insets: zoom in on panels within yellow dotted boxes. Green and red circles are the marks for Fig. 3a and Fig. 3b, respectively.

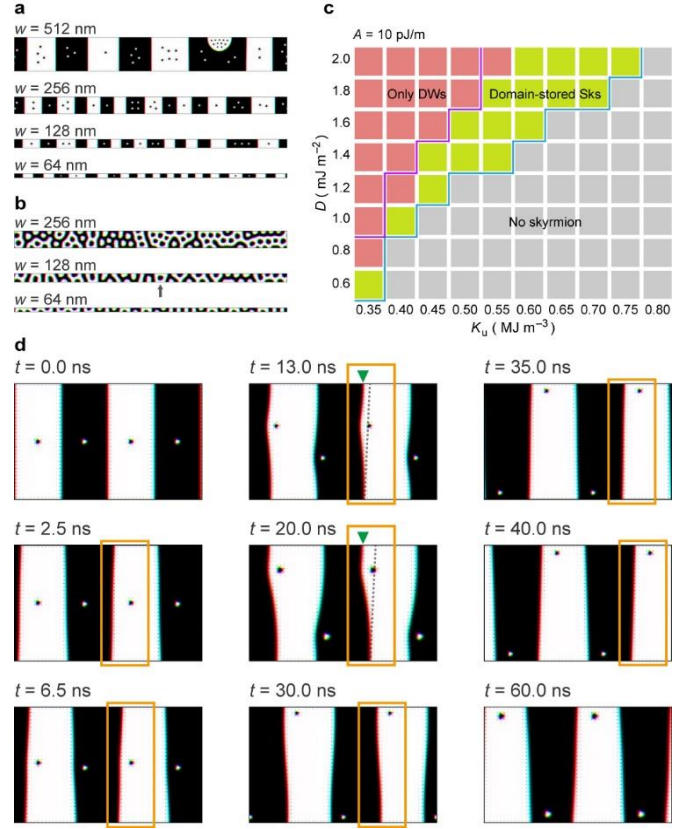

**Fig. 3 | Formation and dynamics of domain-stored skyrmion structure in nanowires**. **a**, Evolution of the magnetic structure as a function of $w$ at $A = 10$ pJ/m, $D = 1.2$ mJ m$^{-2}$, and $K_u = 0.45$ MJ m$^{-3}$. **b**, Same as **a**, but only $K_u$ is changed to 0.35 MJ m$^{-3}$. The arrow indicates the skyrmion. The parameters in **a** and **b** correspond to the green and red circle in Fig. 2, respectively. **c**, Colour-coded $D$-$K_u$ phase diagram at $A = 10$ pJ/m. The green, red, and gray panels indicate the regions of domain-stored skyrmion structures, only DWs, no skyrmions in nanowires, respectively. The cyan and magenta lines are the same as in Fig. 2 at $A = 10$ pJ/m. **d**, Time-dependent dynamics of domain-stored skyrmion structure in a part of a nanowire with $w = 512$ nm. $j_{SOT} = 2.0 \times 10^{10}$ A m$^{-2}$ is injected from left to right along the nanowire. The main text describes dynamics within the orange boxes. In the frames of $t = 13.0$ and 20.0 ns, the green triangles are markers to make it easier to see the stopped segment of the DW, and the black dotted lines represent the tilt of the DW in the absence of the skyrmion.

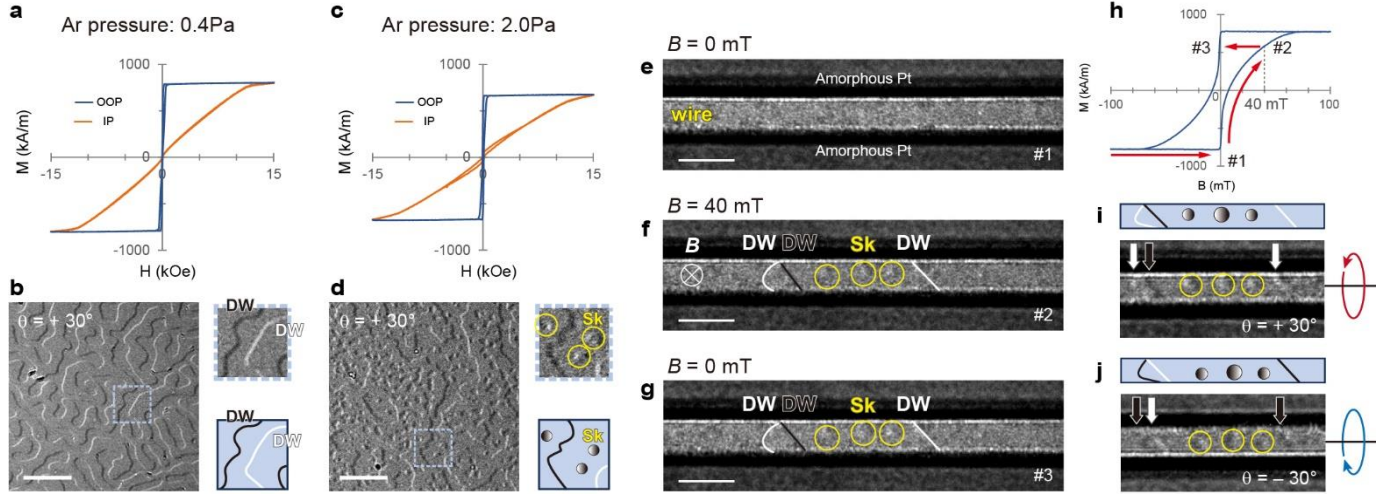

**Fig. 4 | Experiments in Pt/Co/Ta multilayers**. **a**, In-plane (IP, orange) and out-of-plane (OOP, blue) magnetization loops in [Pt(1.3 nm)/Co(1.0 nm)/Ta(0.7 nm)]$_{15}$ thin film by $P_{Ar}$ = 0.4 Pa. The estimated magnetic energies are $K_{eff}$ = 0.50 MJ m$^{-3}$, and $M_s$ = 807 kA/m. **b**, Left: L-TEM image for the as-grown sample in **a** at zero field. The image was obtained at a sample tilt angle of $\theta$ = +30°. Right: Zoom in on the L-TEM image within light blue dotted box and its schematic. The defocus value is $\Delta f$ = −1.8 mm. **c,d**, Same as **a** and **b**, but only $P_{Ar}$ is changed to 2.0 Pa. $K_{eff}$ = 0.48 MJ m$^{-3}$. $M_s$ = 680 kA/m. Sk means the skyrmion. **e-g**, A series of the L-TEM images as a function of out-of-plane magnetic fields in [Pt(1.5 nm)/Co(1.0 nm)/Ta(0.7 nm)]$_{15}$ nanowire with $w$ ≈ 700 nm. The OPP magnetic fields are $B$ ≈ 0 mT after applying $B$ ≈ −2 T (**e**), $B$ ≈ +40 mT (**f**), $B$ ≈ 0 mT after the removal of $B$ ≈ +40 mT (**g**). **h**, The OPP magnetization loop in [Pt(1.5 nm)/Co(1.0 nm)/Ta(0.7 nm)]$_{15}$ thin film. $K_{eff}$ = 0.52 MJ m$^{-3}$. $M_s$ = 784 kA/m. **i,j**, Zoom in on **g**, taking a slightly larger defocus value to enhance the contrasts ($\Delta f$ = −2.8 mm). In **j**, the sample tilts in the inverse direction to **i**. Scale bars, 2 μm (**b,d**), 1 μm (**e–g**).

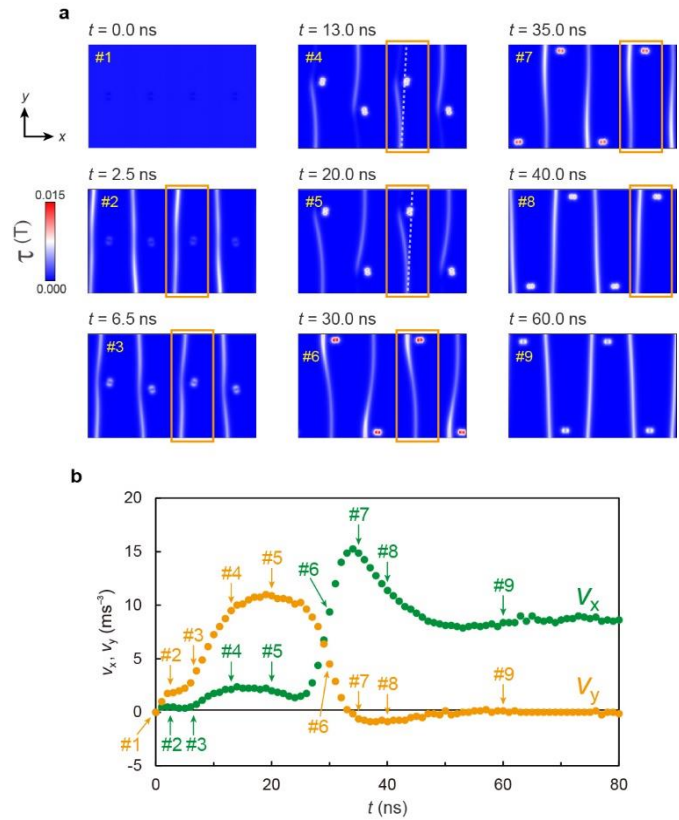

**Fig. 5 | Angular momentum transfers from DWs to skyrmions**. **a**, Density maps of the magnitude of total torque ($\tau$) exerted on magnetic moments. Each panel corresponds to Fig. 3d. The colour scale is shown in the bar. The main text describes dynamics within the orange boxes. In the frames of $t$ = 13.0 and 20.0 ns, the white dotted lines represent the tilt of the DW in the absence of the skyrmion, corresponding to the black dotted lines in Fig. 3d. **b**, Time-dependent skyrmion velocities of the *x*-directional (green) and *y*-directional (orange) components. Time numbers, #1–#9, correspond to those in **a**.

# Supplementary Information

**Table of Contents**

**Supplementary Note 1: *A* dependence of energy and structural boundaries in the *D-K*ᵤ phase diagrams**

In Fig. 2 in the main text, the slopes of the energy boundary lines (magenta lines) and structural boundary lines (cyan lines) lower as *A* decreases. In addition, the lower limit of D for the formation of domain-contained skyrmion structures decreases as *A* decreases. These behaviours can be understood by an equation of the skyrmion stability:

$$\kappa = \frac{\pi D}{4\sqrt{AK_{\text{eff}}}}$$

$$K_{\text{eff}} = K_{\text{u}} - \frac{\mu_0 M_s^2}{2} \quad \text{(for thin films)}$$

where $\mu_0$ is the permeability of vacuum. For $\kappa > 1$, skyrmions are in thermodynamically stable state, whereas for $0 < \kappa < 1$, they are in metastable state. In Fig. 2, the energy boundary lines correspond approximately to $\kappa = 1$. The structural boundary lines correspond to an asymptote to $\kappa = 0$ or correspond to small $\kappa$ values. Realistically, the latter is more appropriate. In the case of the energy boundary lines, when *D* is fixed at a certain value and *A* decreases, $K_{\text{u}}$ increases to reach $\kappa = 1$. When $K_{\text{u}}$ is fixed at a certain value and *A* decreases, *D* quadratically decreases to reach $\kappa = 1$. In other words, as *A* decreases, the position of the energy boundary line shifts toward a quadratic decrease in *D* and an increase in $K_{\text{u}}$. The case of the structural boundary lines can be understood in the same way as the energy boundary lines.

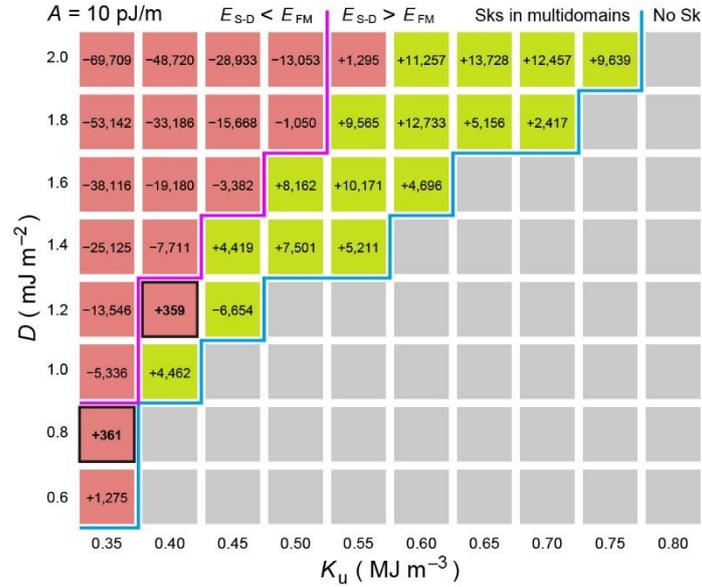

**Supplementary Fig. S1: Energy competition with ferromagnetic states at *A* = 10 pJ/m.** The values in the panel describe energy difference which subtracts the ferromagnetic energy from the domain-contained skyrmion structure energy. Unit, J m⁻³. The meaning of the panel colours is the same as in Fig. 3c in the main text. The panels surrounded by bold lines are magnetic parameter regions where large skyrmions appear, out of the trend of the skyrmion size change. These panels are one to two orders of magnitude smaller than the energy difference values of the surrounding panels.

**Supplementary Table S1:** $K_u$ dependence of the upper wire width for the formation of domain-stored skyrmion structures and the skyrmion sizes, at $A = 10$ pJ/m, $D = 2.0$ mJ m$^{-2}$. As $K_u$ increases, the skyrmion size decreases and the upper limit wire width that can form the structures increases. a: Skyrmions inside the bags (circular domains) are smaller than those outside the bags, due to the repulsion between the periphery of the bags and the inside skyrmions.

| $K_u$ (MJ m$^{-3}$) | 0.60 | 0.65 | 0.70 | 0.70 |
|---|---|---|---|---|
| Skyrmion size (nm) | 16[a]- 42 | 14[a]- 24 | 10 - 18 | 12 - 14 |
| Upper wire width for the formation of domain-stored skyrmion structures (nm) | 64 | 128 | 512 | 2048 |

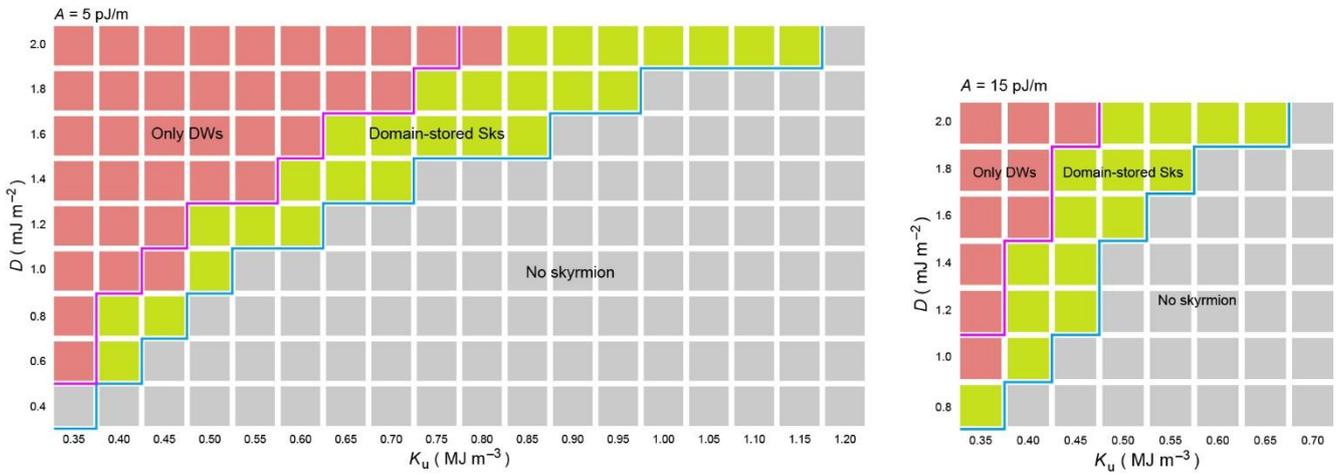

**Supplementary Fig. S2: Colour-coded phase diagrams at $A = 5$ and 15 pJ/m.** Same as Fig. 3c, but the exchange stiffness constant is changed to $A = 5$ (left) and 15 pJ/m (right). The meaning of the panel colours is the same as in Fig. 3c in the main text. The formation regions for domain-stored skyrmion structures in nanowires (green panels) are located only in the $E_{D-S} < E_{FM}$ region at both $A = 5$ and 15 pJ/m, and these patterns are consistent at $A = 10$ pJ/m.

**Supplementary Table S2: List of Pt(3 nm)/[Pt($t_{Pt}$ nm)/Co($t_{Co}$ nm)/Ta($t_{Ta}$ nm)]$_N$/Ta(5 nm)/SiO$_2$.** We prepared 44 samples as below. $H_c$ is coercive field. We estimate $K_u$ according to $K_u = K_{eff} + \mu_0 M_s^2/2$.

| Ar gas pressure [Pa] | Pt thickness [nm] | Co thickness [nm] | Ta thickness [nm] | Number of repetition (N) | $M_s$ [KA/m] | $H_c$ [Oe] | $K_{eff}$ [MJ/m$^3$] | $K_u$ [MJ/m$^3$] |
|---|---|---|---|---|---|---|---|---|
| 0.2 | 0.7 | 0.7 | 0.2 | 10 | 1,182 | 268 | 0.624 | 1,501,935 |
| 0.2 | 0.7 | 0.7 | 0.2 | 10 | 1,088 | 282 | 0.540 | 1,283,960 |
| 0.2 | 0.7 | 0.7 | 0.2 | 7 | 1,143 | 261 | 0.539 | 1,359,791 |
| 0.2 | 0.7 | 0.6 | 0.2 | 7 | 980 | 319 | 0.426 | 1,029,247 |
| 0.2 | 1.4 | 0.7 | 0.4 | 7 | 780 | 77 | 0.348 | 730,149 |
| 0.2 | 0.8 | 0.5 | 0.2 | 8 | 696 | 226 | 0.278 | 582,420 |
| 0.2 | 2.0 | 0.5 | 0.2 | 8 | 728 | small | 0.273 | 605,635 |
| 0.2 | 0.7 | 0.7 | 0.3 | 10 | 856 | small | 0.261 | 721,472 |
| 0.2 | 0.7 | 0.8 | 0.3 | 10 | 944 | small | 0.260 | 819,517 |
| 0.2 | 1.5 | 0.5 | 0.2 | 8 | 700 | small | 0.251 | 558,826 |
| 0.2 | 0.9 | 0.7 | 0.4 | 10 | 682 | small | 0.238 | 530,264 |
| 0.2 | 1.8 | 0.7 | 0.5 | 8 | 545 | small | 0.216 | 402,719 |
| 0.2 | 1.4 | 0.7 | 0.5 | 8 | 477 | small | 0.202 | 344,970 |
| 0.2 | 0.7 | 0.5 | 0.2 | 8 | 642 | 157 | 0.193 | 451,891 |
| 0.2 | 1.5 | 0.7 | 0.5 | 8 | 460 | small | 0.191 | 324,082 |
| 0.2 | 1.5 | 0.5 | 0.2 | 8 | 477 | small | 0.163 | 306,095 |
| 0.2 | 0.8 | 0.7 | 0.4 | 10 | 533 | small | 0.109 | 287,763 |
| 0.2 | 0.7 | 0.7 | 0.4 | 10 | 488 | small | 0.020 | 169,394 |
| 0.2 | 0.5 | 0.8 | 0.5 | 10 | 283 | in-plane | | |
| 0.2 | 0.6 | 0.8 | 0.4 | 10 | 407 | in-plane | | |
| 0.2 | 0.6 | 0.7 | 0.4 | 10 | 271 | in-plane | | |
| 0.2 | 0.7 | 0.7 | 0.4 | 10 | 252 | in-plane | | |
| 0.2 | 1.8 | 0.6 | 0.5 | 8 | 196 | in-plane | | |
| 0.4 | 3.0 | 1.0 | 1.9 | 3 | 781 | 58 | 0.430 | 812,800 |
| 0.4 | 1.1 | 1.0 | 0.8 | 3 | 1,040 | 55 | 0.426 | 1,105,989 |
| 0.4 | 1.2 | 1.0 | 0.8 | 3 | 1,040 | 58 | 0.390 | 1,069,589 |
| 0.4 | 1.2 | 1.0 | 0.8 | 3 | 860 | | 0.361 | 825,904 |
| 0.4 | 1.2 | 1.0 | 0.8 | 3 | 913 | 58 | 0.342 | 866,122 |
| 0.4 | 1.3 | 1.0 | 0.7 | 15 | 807 | | 0.495 | 903,964 |
| 0.4 | 1.4 | 1.0 | 0.6 | 15 | 892 | | 0.519 | 1,018,851 |
| 2.0 | 1.7 | 1.1 | 0.7 | 15 | 957 | 80 | 0.635 | 1,210,893 |
| 2.0 | 1.7 | 1.0 | 0.7 | 15 | 912 | 95 | 0.632 | 1,155,072 |
| 2.0 | 1.5 | 0.9 | 0.7 | 15 | 733 | 70 | 0.516 | 853,987 |
| 2.0 | 1.5 | 1.0 | 0.7 | 15 | 784 | 75 | 0.516 | 902,464 |
| 2.0 | 1.7 | 0.9 | 0.7 | 15 | 674 | 84 | 0.476 | 761,611 |
| 2.0 | 1.5 | 0.8 | 0.6 | 15 | 683 | 75 | 0.452 | 744,908 |
| 2.0 | 1.5 | 0.7 | 0.5 | 15 | 676 | 79 | 0.431 | 717,738 |
| 2.0 | 1.3 | 1.0 | 0.7 | 15 | 848 | 211 | 0.416 | 867,770 |
| 2.0 | 1.5 | 0.8 | 0.7 | 15 | 654 | 95 | 0.410 | 678,800 |
| 2.0 | 1.5 | 0.8 | 0.6 | 15 | 623 | 46 | 0.397 | 640,408 |
| 2.0 | 1.3 | 1.0 | 0.7 | 10 | 713 | 67 | 0.392 | 711,568 |
| 2.0 | 1.5 | 0.8 | 0.7 | 15 | 541 | 48 | 0.344 | 527,702 |
| 2.0 | 1.5 | 0.7 | 0.5 | 15 | 539 | | 0.337 | 519,415 |
| 2.0 | 1.3 | 1.0 | 0.7 | 15 | 680 | | 0.482 | 772,009 |

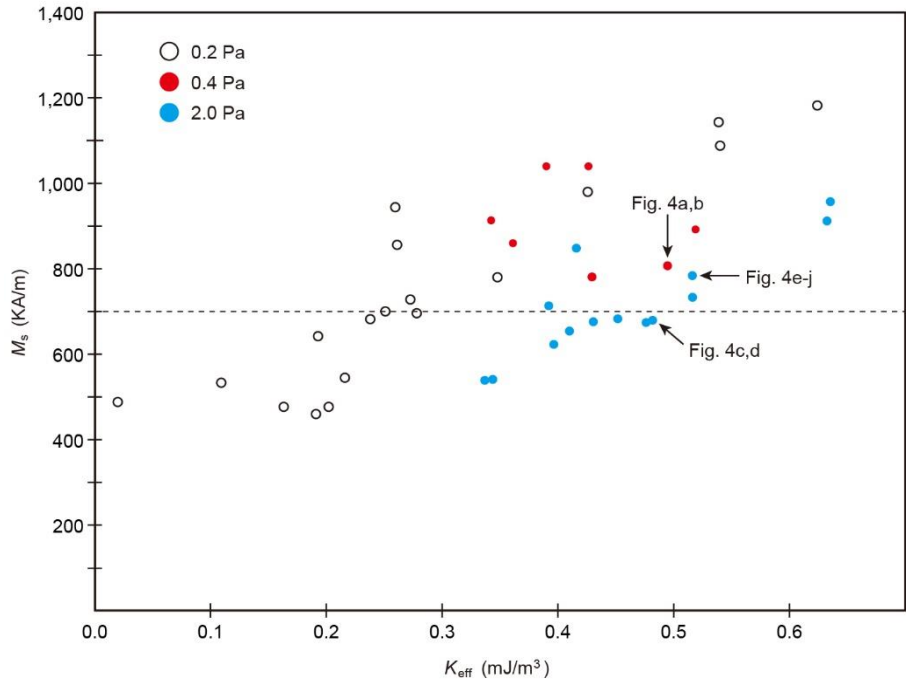

**Supplementary Fig. S3: Plots of $M_s$ vs. $K_{eff}$ for our Pt/Co/Ta samples.** Open black, filled red, and filled light blue circles represent samples prepared at $P_{Ar} = 0.2$, 0.4, and 2.0 Pa, respectively. The dotted line at $M_s = 700$ KA/m is at half the magnetization of bulk Co, which is the value we used in micromagnetic simulations. We aimed for large effective perpendicular magnetic anisotropy $K_{eff}$ while paying attention to the saturation magnetization $M_s$ value and the number of repetition layers $N$ because there is a trade-off between increasing L-TEM contrast intensity and suppressing dipolar interaction. The samples displayed in Fig. 4 have relatively low $M_s$ and relatively high $K_{eff}$ among the samples that we prepared.

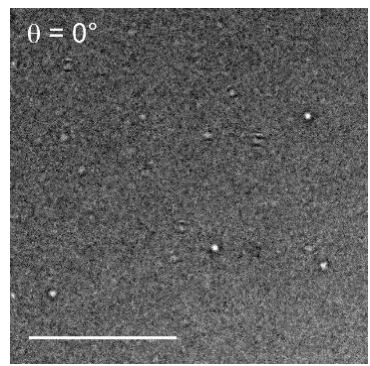

**Supplementary Fig. S4: L-TEM image at $\theta = 0°$ in [Pt(1.3 nm)/Co(1.0 nm)/Ta(0.7 nm)]$_{15}$ thin film of 0.4-sample.** The sample prepared by magnetron sputtering at $P_{Ar} = 0.4$ Pa is identical to that in Fig. 4b in the main text but only the sample tilt angle is changed to $\theta = 0°$. There is no contrast originating from magnetization. This result together with the result in Fig. 4b, indicates that Néel-type DWs are dominant in this sample. The bright spots are pores in the $Si_3N_4$ membrane substrate. Scale bar, 2 μm.

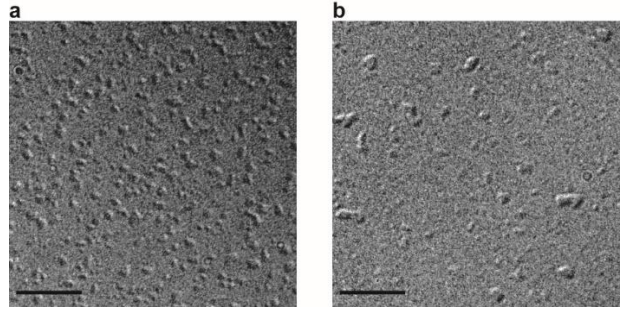

**Supplementary Fig. S5: L-TEM images for dense and sparse skyrmion states in thin films. a,b**, L-TEM images obtained at zero fields after applying $B = 2$ T for [Pt(1.3 nm)/Co(1.0 nm)/Ta(0.7 nm)]$_{15}$ (**a**) and [Pt(1.5 nm)/Co(1.0 nm)/Ta(0.7 nm)]$_{15}$ (**b**). These samples were grown by magnetron sputtering at an Ar gas pressure of $P_{Ar} = 2.0$ Pa. In **a**, and **b**, dense and sparse skyrmion states appear, respectively, suggesting stable and metastable states, respectively. We fabricated the latter sample into nanowires. Scale bars, 2 μm.

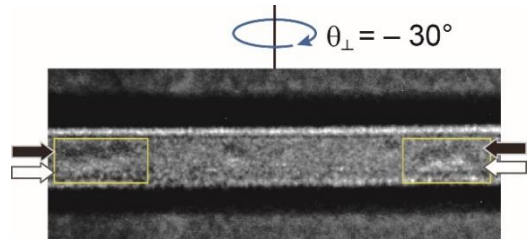

**Supplementary Fig. S6:** L-TEM image in nanowire with width of $w = 700$ nm. The sample tilt axis differs from Fig. 4e-g,i,j by 90 degrees, which is the imaging condition that emphasizes DWs along the longitudinal direction of the wire. The L-TEM image shows a few DWs nearly parallel to the longitudinal direction (bright and dark contrasts within the yellow boxes), indicating that the wire width is wider than adequate.

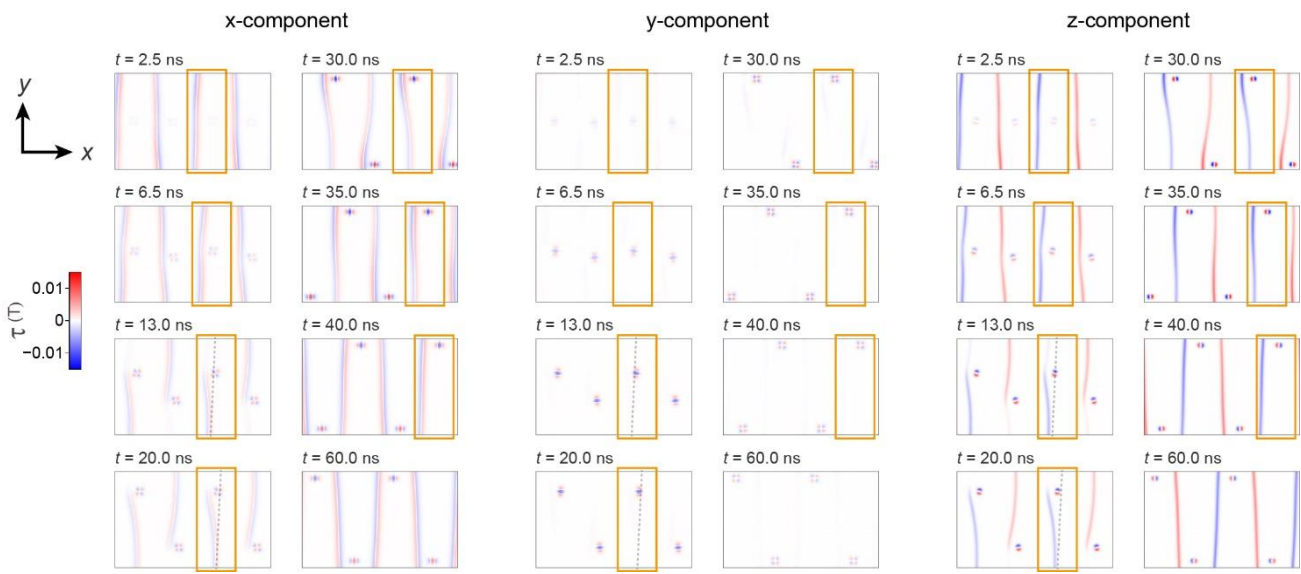

**Supplementary Fig. S7:** Density maps of torque resolving Fig. 5a into $x$-, $y$-, and $z$-components. The colour scale is shown in the bar.

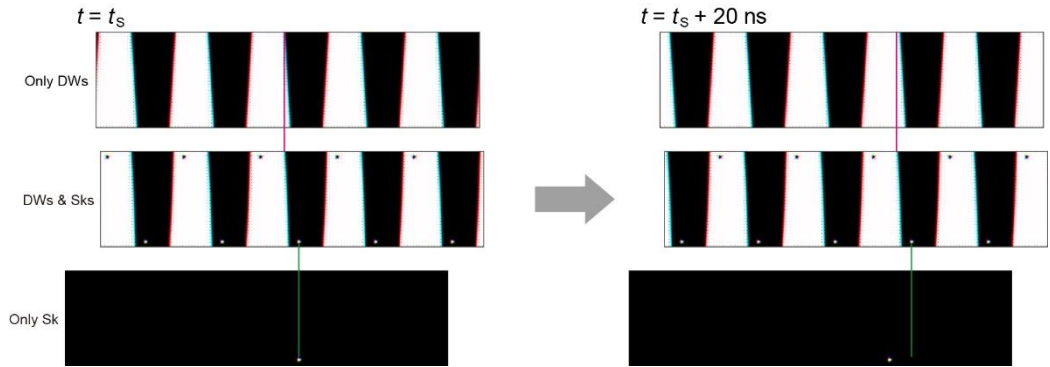

**Supplementary Fig. S8: Difference in steady-state motion of DWs and skyrmions dependent on the systems.** Upper row: only DWs, middle row: coexistence of DWs and skyrmions, lower row: only isolated skyrmion. Left: Magnetization distribution maps at any time $t_S$ in the steady state. Right: Magnetization distribution maps after 20 ns ($t_S$ + 20 ns). Magenta lines are the DW position guidelines. The green lines are the guideline for the skyrmion position. The DW (skyrmion) motion in the coexisting system of DWs and skyrmions is slower (faster) than in the absence of skyrmions (DWs), indicating a steady leakage of angular momentum from DWs to skyrmions.

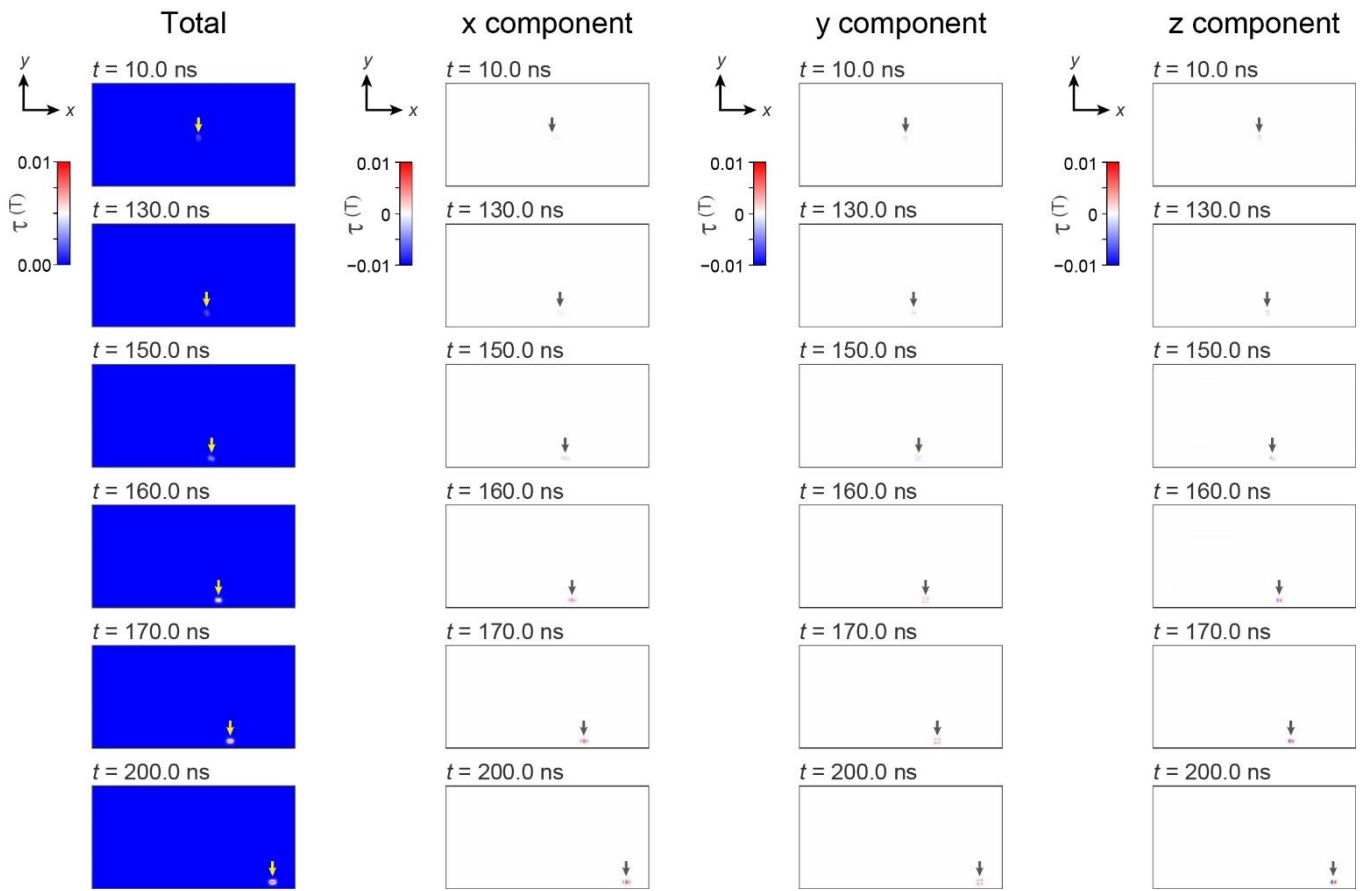

**Supplementary Fig. S9: Density maps of total torque and their $x$-, $y$-, and $z$-components in the nanowire with an isolated skyrmion.** The colour scale is shown in the bars. The arrows indicate the position of the skyrmion. The change in torque of the $x$-, $y$-, and $z$-components is consistent with the change in total torque. When the skyrmion reaches the wire edge and move along the $x$-direction, the density of every torque increases.